\documentclass[prl,nofootinbib,twocolumn]{revtex4}
\usepackage{graphicx}
\usepackage{amsmath}
\usepackage{amsfonts}
\usepackage{graphics}
\usepackage{bm}
\usepackage{float}

\begin{document}

\title{Cosmic Sparks from Superconducting Strings}

\author{Tanmay Vachaspati}
\affiliation{Institute for Advanced Study, Princeton, NJ 08540.\\
CERCA, Physics Department, Case Western Reserve University,\\
10900 Euclid Avenue, Cleveland, OH 44106-7079.}

\begin{abstract}
\noindent
We investigate cosmic sparks from cusps on superconducting 
cosmic strings in light of the recently discovered millisecond 
radio burst by Lorimer {\it et al} \cite{Lorimer:2007qn}. 
We find that the observed duration, fluence, spectrum, and
event rate can be reasonably explained by Grand Unification 
scale superconducting cosmic strings that carry currents 
$\sim 10^5$ GeV. The superconducting string model predicts 
an event rate that falls off only as $S^{-1/2}$, where
$S$ is the energy flux, and hence predicts a population 
of very bright bursts. Other surveys, with different 
observational parameters, are shown to impose tight 
constraints on the superconducting string model.
\end{abstract}

\maketitle

The discovery of a radio burst (``spark'') was recently 
reported by Lorimer {\it et al} \cite{Lorimer:2007qn} in
the Parkes survey and 
analysis of the dispersion measure suggests that the source 
is of cosmological origin. A thorough examination of the 
observation has supported this conclusion \cite{Kulkarnietal}. 
Conventional astrophysical sources are not known and neither 
has a host galaxy for the event been identified. 

If more cosmic sparks are observed and found to be extra-galactic,
it would indicate an exotic cosmological process.
Superconducting cosmic strings are a possible exotic source of 
electromagnetic phenomena in the universe and arise quite naturally 
in particle physics models \cite{Witten:1984eb}, though less
so in string theory \cite{Polchinski:2004ia}. Earlier work
on cosmological signatures of superconducting cosmic strings
has primarily foucussed on high energy emission in the 
form of particles and/or gamma rays 
\cite{Witten:1984eb,Babul:1986wd,Berezinsky:2001cp} or
synchrotron emission \cite{Chudnovsky:1986hc}. 
In this paper we argue that radio emission may be a good
way to look for superconducting strings as they can cause 
observable sparks similar to the one seen by Lorimer {\it et al}. 
Even if further observations discover astrophysical sources for 
observed sparks, the prediction that superconducting cosmic strings 
produce radio sparks that are potentially observable is important 
from the particle physics viewpoint, since their detection or 
absence may be used to constrain various fundamental models.

\begin{figure}
\scalebox{0.80}{\includegraphics{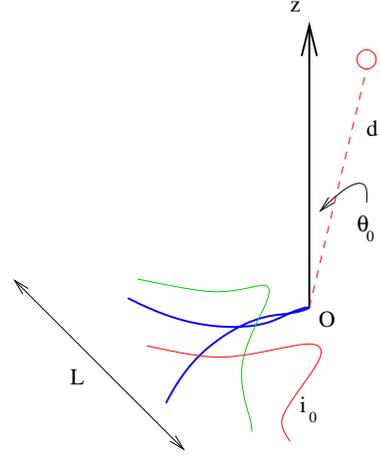}}
\caption{String segment with current $i_0$ at 3 different times, 
with a cusp at O, with velocity along $z$. The size of the 
curved section of string is $L$. The observer is at an
angle $\theta_0$ from the z-axis at a distance $d$.}
\label{systemfig}
\end{figure}

Superconducting cosmic strings can be viewed as elastic, 
current-carrying wires, distributed in the cosmos as 
closed loops and infinitely long Brownian curves. The mass per 
unit length of a string will be denoted by $\mu = \eta^2$ and the 
current by $i_0$. 
Current-carrying strings oscillate under their own tension and 
radiate electromagnetically ({\it e.g.} see \cite{VilenkinShellard}). 
The radiation is very strong from events such as ``cusps'' which
are points on an idealized (zero thickness, no current) string 
that reach the speed of light for a brief instant
\cite{Vilenkin:1986zz,BlancoPillado:2000xy,BlancoPillado:2000ep}.
In a more
realistic setting, the cusp gets cut-off due to the thickness
of the string and due to the backreaction of the current and radiation.
Nonetheless the radiation is very strong from localized regions 
(``quasi-cusps'') even in the realistic string case. The scenario 
we envision is that a curved section of string (or loop) of length 
$L$ develops a cusp and beams electromagnetic radiation in direction 
${\hat z}$. The observer is located at some large distance, 
$d$, from the location of the cusp and slightly off the $z$ axis, 
at an angle $\theta_0$ (Fig.~\ref{systemfig}).

The energy emitted from a cusp at angular frequency 
$\omega = 2\pi \nu$, at angle $\theta$ from the beam 
direction, and observed at distance $d$
({\it i.e.} the fluence) is given by 
\begin{equation}
F \equiv  \frac{1}{d^2} \frac{d^2 E_\omega}{d\omega d\Omega} 
  \sim b i_0^2 \frac{L^2}{d^2}
           e^{-a \omega L \theta^3} \ \ \ {\rm if} \ \ \ 
a \omega L \theta^3 > 1
\label{Eradiation}
\end{equation}
where $a$ and $b$ are constants that depends on the shape of 
the cusp and we will take $a \sim 1$, $b \sim 1$. (We also
work in natural units throughout, so $\hbar=1=c$.) This 
result is most easily obtained by expanding the result in
Ref.~\cite{BlancoPillado:2000xy} for large $\omega L\theta^3$. 
If, instead, we have $\omega L\theta^3 < 1$, the emitted energy 
falls off $\propto \omega^{-2/3}$.

The total energy emitted from a loop in electromagnetic and
gravitational radiation in one time period is 
$
E_{\rm total, loop} \sim (\kappa e i_0 \eta  + \Gamma G \mu^2 ) L
                    \equiv \gamma \mu L
$
where $e \approx 0.3$ is the charge of the current carriers,
$G$ Newton's gravitational constant,
and $\kappa \sim 30$, $\Gamma \sim 100$ are numerical factor 
determined by averaging numerically derived radiation rates 
from a variety of loops. 

The temporal width of the emission seen from the cusp is given 
by the duration for which the observer is in the beam. The emission 
from the cusp is beamed but the beam is changing direction at a 
rate ${\dot \theta} \sim 1/L$. So the beam changes its direction 
by $\theta_0$ in a time $L\theta_0$. 
Since, for large angles from the beam, the emission gets cutoff 
exponentially fast (Eq.~(\ref{Eradiation})), we have
$\theta_0 \sim (\omega L)^{-1/3}$, and
\begin{equation}
\Delta t \sim \omega^{-1/3} L^{2/3}
\label{duration}
\end{equation}

The event rate depends on the number density of loops and 
is derived from numerical simulations of non-superconducting
strings \cite{VilenkinShellard}. Note that the results for
non-superconducting strings should also apply in the case 
that the current on the string is much smaller than the
symmetry breaking scale $\eta$. The simulations ignore 
radiative effects and find that the number density of loops 
of length between $l_0$ and $l_0+dl_0$ is
$
dn_{l_0} \sim A {dl_0}/{(l_0^2 t^2)}
$
where $A \sim 10^2$ 
\cite{Martins:2005es,Vanchurin:2005pa,Ringeval:2005kr}
(also see \cite{footnote}).
Due to radiation, the length $l$ of a loop decreases 
with time, $l(t) = l_0 - \gamma (t-t_i)$, where 
$t_i$ is the time when the loop was born. So the number density of 
loops of size $l$ at time $t$ is
\begin{equation}
dn_l (t) \sim 
\frac{A dl}{(l+\gamma  t)^2 t^2}
\label{numberdensity}
\end{equation}
where the radiative shrinking has been included, and we have
assumed that the loops were all born at some very early time 
so that $t \gg t_i$.

The spark observed by Lorimer {\it et al} was in the frequency 
interval (1.28,1.52) GHz and the central frequency is
$\nu_0 = 1.4 ~ {\rm GHz}$.
Based on the dispersion measure of the event, the observed 
event is constrained to lie within redshift of $0.1-0.3$ and,
for our estimates, we will assume that the event was located 
at $z_0=0.3$, or at a comoving distance $\sim 1$ Gpc.
The observed energy flux is
\begin{equation}
S_{\rm obs} = 30 ~ {\rm Jy} = 3 \times 10^{-22} ~ {\rm \frac{ergs}{cm^2-s-Hz}}
\label{obsfluence}
\end{equation}
and the duration of the event has an upper bound
$
\Delta \tau \lesssim \Delta \tau_0 = 5 ~ {\rm ms} 
$

The observed pulse width cannot be used to estimate the intrinsic
duration of the event because of scattering by the turbulent
inter-galactic medium (IGM) \cite{Lorimer:2007qn,Kulkarnietal}. 
The oberved time width, $\Delta t_{\rm obs}$, is a sum in
quadratures of the intrinsic time width modified by 
cosmological redshift and the width due to scattering 
in the IGM:
$
\Delta t_{\rm obs} = \left [ (\Delta \tau_{\rm obs} )^2 +
                 (\Delta t_{\rm emit})^2 \right ]^{1/2}
$
with
\begin{equation}
\Delta t_{\rm emit} = (1+z) \omega_{\rm emit}^{-1/3} l^{2/3}
             = \omega_0^{-1/3} (1+z)^{2/3} l^{2/3}
\label{intdt}
\end{equation}
The scattering time width at frequency $\nu$ from an
event at redshift $z$ depends on the location of the
scattering centers. For the case of a scattering screen
close to the source, the width is 
\cite{LeeJokipii1976,Kulkarnietal}
\begin{equation}
\Delta \tau_{\rm obs} (\nu, z)
= \Delta \tau_0 \left ( \frac{1+z}{1+z_0} \right )^{\beta +1} 
     \left ( \frac{\nu}{\nu_0} \right ) ^\beta
\end{equation}
where $\Delta \tau_0 = 5$ ms, $\nu_0 = 1.4$ GHz, $z_0=0.3$,
and $\beta = -4.8$. For comparison with the observed
spark in the Parkes survey, we will use $\nu=\nu_0$, but 
if comparing to other surveys it will be necessary to 
insert the appropriate observational frequency.

The observed power law fall off $\propto \nu^{-4}$ in the 
observational frequency band can be fit by an exponential,
and since the fluence is $S_{\rm obs} \Delta t_{\rm obs}$,
\begin{eqnarray}
F_{\rm obs} 
&=& \frac{S_{\rm obs}}{2\pi} \Delta \tau_0
\sqrt{1+ \left ( \frac{\Delta t_0}{\Delta \tau_0} \right ) ^2 }
e^{-(\nu -\nu_0) /\nu_c} \nonumber \\
&\approx& 10^{-23} e^{-4 \nu/\nu_0} ~ {\rm \frac{ergs}{cm^2-Hz}}
\label{obsF}
\end{eqnarray}

To get an idea of the parameters needed of the superconducting
string model, we first fit the observed spectrum ignoring redshift
factors (which are small since $z_0 < 0.3$). Using 
Eq.~(\ref{Eradiation}) we find
\begin{equation}
\frac{d\ln F}{d\ln \omega} = - a \omega L \theta^3 = -4
\label{expvalue}
\end{equation}

The observation does not constrain the intrinsic duration 
of the event. However, if the intrinsic duration was $1$ ms, 
we can equate the duration of the cusp event (Eqs.~(\ref{duration}))
to the observed duration 
to get
$\omega_0^{-1/3} L^{2/3} = \Delta t_0 = 1 ~ {\rm ms}$,
where $\omega_0=2\pi \nu_0 \sim 10^{10} ~ {\rm s}^{-1}$.
This gives
$L  = \omega_0^{1/2} (\Delta t )^{3/2} \sim 3 ~{\rm s} 
                               =   10^{11} ~ {\rm cm}$.
If the intrinsic duration were smaller than 1 ms, the 
length would be even smaller. This illustrates 
that we are considering signatures of loops that are
very small on a cosmological scale.

The energy flux from the cusp event is found from
Eqs.~(\ref{Eradiation}) and (\ref{expvalue}). 
Equating to the observed value, Eq.~(\ref{obsF}), gives
\begin{equation}
i_0 \sim 
10^5 \left ( \frac{10^{11} {\rm cm}}{L} \right )
     \left ( \frac{d}{1 ~{\rm Gpc}} \right ) ~ {\rm GeV}
\end{equation}
where we have used 
$1 ~{\rm ergs/(cm^2-Hz)} = 1 ~{\rm GeV}^2$. Note also that the 
dynamics of the string will be dominated by the tension if 
$\eta \gg 10^6$ GeV. Hence it is a valid approximation to ignore 
the current when discussing the dynamics of the string network.

To estimate the rate of observed-sparker-like events,
we use Eq.~(\ref{numberdensity}). 
Then the event rate, denoted ${\dot N}$, due to cusps on 
loops of size between $l$ and $l+dl$, beamed within a solid 
angle $d\Omega$ from us is
\begin{equation}
d{\dot N} \sim \frac{f_c}{l} ~
  \frac{A dl}{(l+\gamma t_0 )^2 t_0^2} ~ 
          \frac{d\Omega}{4\pi} ~ dV
\label{rate}
\end{equation}
where the factors account for $f_c$ cusps per loop oscillation,
the number density of loops, the beaming angle constraint, and
the spatial volume. For Grand Unification scale strings,
$\eta \sim 10^{14}$ GeV, we find 
$\gamma t_0 \sim 10^{9} ~ {\rm s} \gg l \sim 1 ~{\rm s}$, 
and we can rewrite (\ref{rate}) as
\begin{equation}
d{\dot N} \sim C \times 
 \frac{dl}{l} \times \sin\theta d\theta \times D^2 dD
\label{erate}
\end{equation}
where $D \equiv H_0 D_c$ is the comoving distance to 
the loop in Hubble units, and
$C \equiv {2\pi A f_c H_0}/{\gamma^2}
               \sim  10^7 f_c \ {\rm day}^{-1}$,
for $\gamma = 10^{-8}$.
Now that we know the rate of events as a function of
$l$, $\theta$ and $D_c$, we can also evaluate the 
flux density in these terms 
\begin{equation}
S(l,\theta, D_c) = 
\frac{2 \pi i_0^2 l^2}{D_c^2 (\Delta t)_{\rm obs}}
             e^{-\omega_0 (1+z) l \theta^3} ~ ,
\ \ \omega_0 (1+z) l \theta^3 > 1
\end{equation}
and we assume a ``top hat'' cut-off at small angles,
$S(l,\theta, D_c) = 
2 \pi e^{-1} i_0^2 l^2/D_c^2 (\Delta t)_{\rm obs}$
if $\omega_0 (1+z) l \theta^3 \le 1$.

The event rate relevant to a survey is
\begin{equation}
{\dot N} (  > S ) = C \int_{\cal V} 
           \frac{dl}{l} ~ D^2 dD ~ \sin\theta d\theta 
\end{equation}
Here the integration volume ${\cal V}$ is constrained by the
requirement that the flux be greater than $S$ and that 
the observed duration of the event, $(\Delta t)_{\rm obs}$,
be in the range in which the search was carried out.  
In our case, $(\Delta t)_{\rm obs}$ is in the
interval (1 ms, 1 s) to coincide with the search parameters 
in Ref.~\cite{Lorimer:2007qn}, and we will assume a standard
flat cosmology. After integrating over $\theta$ by hand, 
the remaining integrals were evaluated numerically, giving
\begin{equation}
{\dot N} (  > 30 ~{\rm Jy} ) \sim 10^0 f_c ~ {\rm day}^{-1}
\end{equation}
whereas the single observed event gives a 99\% double-sided
confidence level estimate between $2$ to $3\times 10^3$ 
events per day \cite{Kulkarnietal}. (Based on the dynamics 
of smooth loops, one expects $f_c \sim 1$.) The absence of 
fainter events, but still above the threshold of 0.3 Jy, is 
not significant since the prediction is only around 10 events 
above 0.3 Jy, and also the reach of the Parkes survey at threshold 
is less by a factor $\sim 10$ than at 30 Jy \cite{Kulkarnietal}.
This shows that the superconducting string model can give an 
event rate consistent with the Lorimer et al observations. 
Further, it is possible to
show that for fixed observational frequency and 
duration range, ${\dot N} (> S) \propto S^{-1/2}$ for 
large $S$. This is quite different from the $S^{-3/2}$ fall 
off expected from uniformly distributed, identical sources. 
The slow fall off implies that there
may be a population of very strong sparks {\it e.g.} $\sim 1$
per year above 1 MJy.

\begin{figure}
\includegraphics[width=2.5in,angle=-90]{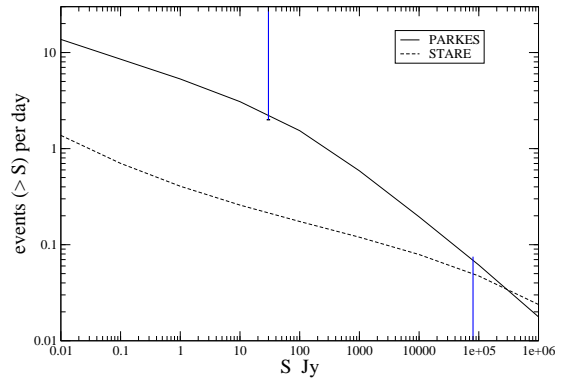}
\caption{Event rates for PARKES and STARE survey parameters.
The vertical line at 30 Jy corresponds to the inferred rate
from the PARKES survey and at 80000 Jy shows the constraint
from the STARE survey.
}
\label{eratefig}
\end{figure}

A more thorough analysis, taking into account the constraints 
imposed by different surveys, may impose further limits on 
the model. For example, the STARE survey \cite{Katz:2003et}
places an upper bound of $7.5\times 10^{-2}$ events per
day \cite{Kulkarnietal} for $S > 80$ kJy at an observational 
frequency of 611 MHz and durations ranging from 125 ms to a 
few minutes. With these observational parameters and the 
superconducting string model parameters used above, the model 
prediction is $\sim 6\times 10^{-2}$ events per day. This 
suggests that useful constraints on the string model may 
be placed by presently existing data (see Fig.~\ref{eratefig}). 
In particular, a value of $\gamma$ significantly less than 
$10^{-8}$ is already ruled out.

While we expect a quasi-cusp to repeat every oscillation, we 
do not expect the beaming direction to remain the same in 
every repetition. 
The angular momentum carried off by the beamed radiation is 
$\sim E_{\rm beam} L$ and the moment of inertia of
the string in the cusp region is $\sim (\mu\sigma ) L^2$,
where $\sigma \sim (\omega_0 L)^{-1/3} L$ is the 
length of string in the cusp region relevant to
the observed radiation frequency. The angular velocity 
due to radiation backreaction is 
${\dot \theta} \sim E_{\rm beam}L/\mu \sigma L^2$. 
Since the time interval between two cusp events is
$\sim L$, we estimate the change in the beaming angle as
\begin{equation}
\Delta \theta \sim 
 \frac{E_{\rm beam}L}{\mu  \sigma L^2} L 
  \sim \frac{\kappa e i_0}{\eta} (\omega_0 L)^{1/3} 
       \sim 2 \times 10^{-4}
\end{equation}
which corresponds to a change in the observed flux 
$
| \Delta \ln F | \sim |a\omega_0 L \theta_0^3|
\times 3 \Delta \theta/\theta_0 \sim 5
$
Hence the beamed fluence may get reduced by a factor 
$e^{-5} \sim 10^{-2}$.
This estimate suggests that sparks due to consecutive
cusps on strings would be beamed in only slightly different 
directions but, since the fluence is highly sensitive to the
beaming direction (Eq.~(\ref{Eradiation})), the repeated
event may or may not be observed, depending on the precise 
cusp shape and other parameters. In the 20 days that 
Lorimer et al searched for bursts, the particular loop
would have on order $10^6$ cusps, and since the solid
angle of the beam is order $10^{-6}$, we can expect on
order one event in those 20 days. Also, cusps that occur
much later, will occur with different parameters since
the length and shape of the loop are likely to change
due to backreaction.

To model the observations, we have taken strings whose 
tension scale is $\eta \sim 10^{14}$ GeV (so that 
$\gamma \sim 10^{-8}$)
but with currents of only $\sim 10^5$ GeV. The small current
compared to the tension scale can have a natural origin in
terms of scattering of counter-propagating particles on
the string \cite{Barr:1987xm,Barr:1987ij} and/or the modified 
dispersion relation due to background magnetic fields 
\cite{Ferrer:2006ne}.

We have considered only the direct emission of 1.4 GHz
radiation from the cusp. This corresponds to a very high
harmonic emitted from the oscillating loop. The fundamental
frequency is given by $L^{-1} \sim 1$ Hz, and 1 GHz 
emission corresponds to the $10^9$th harmonic. What happens 
to emission at lower frequencies?
Since the cusp event is not expected to be in a host galaxy,
it is surrounded only by the IGM with free electron number
density $n_e \sim 10^{-7} ~{\rm cm}^{-3}$ and 
plasma frequency $\omega_p = (4\pi n_e e^2/m_e)^{1/2} \sim 
30 ~ {\rm s}^{-1}$ or $\nu_p \sim 5 {\rm Hz}$. The energy 
emitted from the cusp at $\nu \lesssim 5$ Hz cannot propagate 
in the IGM and must push it around, creating shocks, 
fireballs, and possibly gamma ray bursts 
\cite{Ostriker:1986xc,Babul:1986wd,Berezinsky:2001cp}. 
Assuming that all the very low frequency emission gets
converted into gamma rays due to plasma effects, 
the power emitted is $\sim i_0^2 = 10^{10} {\rm GeV}^2$. 
For gamma rays at $1 {\rm GeV}$ and for loops at a 
distance of $1 {\rm Gpc}$, this gives a photon flux
of $10^{-22} {\rm cm^{-2} s^{-1}}$ which is far below
the threshold, $\sim 10^{-8} {\rm cm^{-2} s^{-1}}$ of the 
Third Interplanetary Network.

The electromagnetic emission from strings also distorts 
the cosmic microwave background spectrum and this effect
has been used to constrain superconducting strings
\cite{Ostriker:1986xc,Sanchez:1991jv}. 
Our choice of
parameters, $G\mu \ll 10^{-6}$, is within these constraints. 
It is worth pointing out that the earlier work
on gamma ray bursts from cusp events primarily focussed
on superconducting strings placed in the galactic 
environment. More detailed investigation is needed to
determine if gamma ray bursts are expected to accompany
radio sparks occurring in the IGM.

The superconducting string model may be tested in a variety 
of ways. Gravitational effects of strings with 
$G\mu \sim 10^{-10}$
will be weak and may not be within forseeable detection 
capabilities, except possibly for gravitational wave
bursts from cusps 
\cite{Damour:2000wa}. 
A promising possibility is to look for signatures of 
particle emission, such as positrons \cite{Ferrer:2005xva}
or other decaying particles produced where the current on
the string quenches. These particles would give a distinctive 
feature in the emission from the vicinity of the spark.
In addition to the observed sparker like events, there 
should be rarer events where we are even closer 
to the beam, such that $\omega L \theta^3 < 1$. Then the 
spectrum will not decay exponentially, and the characteristics 
of the event should be quite different. Also, kinks on 
superconducting strings will radiate in unusual ``fan-like''
patterns \cite{Garfinkle:1987yw}. Most immediately, it is 
necessary to find more sparks and check if they are associated 
with host galaxies since supermassive strings of the kind we 
have considered are expected to roam outside of galaxies.

I am grateful to Mark Hindmarsh, Mario Juric, Shri Kulkarni, 
Ken Olum, Jerry
Ostriker, Edward Witten and Zheng Zheng for discussions. This 
work was supported in part by the U.S. Department of Energy 
and NASA at Case Western Reserve University.


\begin{thebibliography}{999}

\bibitem{Lorimer:2007qn}
  D.~R.~Lorimer, M.~Bailes, M.~A.~McLaughlin, D.~J.~Narkevic and F.~Crawford,
  arXiv:0709.4301 [astro-ph].

\bibitem{Kulkarnietal}
S.R.~Kulkarni, E.O.~Ofek, J.D.~Neill, M.~Juric and Z.~Zheng,
``Giant Sparks at Cosmological Distances,'' unpublished (2007).

\bibitem{Witten:1984eb}
  E.~Witten,
  Nucl.\ Phys.\  B {\bf 249}, 557 (1985).

\bibitem{Polchinski:2004ia}
  J.~Polchinski,
  arXiv:hep-th/0412244.

\bibitem{Babul:1986wd}
  A.~Babul, B.~Paczynski and D.~Spergel,
  Ap.\ J.\ Lett.\ {\bf 316}, L49 (1987)

\bibitem{Berezinsky:2001cp}
  V.~Berezinsky, B.~Hnatyk and A.~Vilenkin,
  Phys.\ Rev.\  D {\bf 64}, 043004 (2001)

\bibitem{Chudnovsky:1986hc}
  E.~M.~Chudnovsky, G.~B.~Field, D.~N.~Spergel and A.~Vilenkin,
  Phys.\ Rev.\  D {\bf 34}, 944 (1986).

\bibitem{VilenkinShellard}
``Cosmic Strings and Other Topological Defects,''
A.~Vilenkin and E.P.S.~Shellard,
C.U.P., 1994.

\bibitem{Vilenkin:1986zz}
  A.~Vilenkin and T.~Vachaspati,
  Phys.\ Rev.\ Lett.\  {\bf 58}, 1041 (1987).

\bibitem{BlancoPillado:2000xy}
  J.~J.~Blanco-Pillado and K.~D.~Olum,
  Nucl.\ Phys.\  B {\bf 599}, 435 (2001)

\bibitem{BlancoPillado:2000ep}
  J.~J.~Blanco-Pillado, K.~D.~Olum and A.~Vilenkin,
  Phys.\ Rev.\  D {\bf 63}, 103513 (2001)

\bibitem{Martins:2005es}
  C.~J.~A.~Martins and E.~P.~S.~Shellard,
  Phys.\ Rev.\  D {\bf 73}, 043515 (2006)

\bibitem{Vanchurin:2005pa}
  V.~Vanchurin, K.~D.~Olum and A.~Vilenkin,
  Phys.\ Rev.\  D {\bf 74}, 063527 (2006)

\bibitem{Ringeval:2005kr}
  C.~Ringeval, M.~Sakellariadou and F.~Bouchet,
  JCAP {\bf 0702}, 023 (2007)

\bibitem{footnote}
The loop distribution is still under study {\it e.g.}
\cite{Dubath:2007mf}. In \cite{Vincent:1997cx}
the authors find that loops will only be of microscopic
size, in which case radio emission is only possible
from cusps on long strings. In \cite{Rocha:2007ni}, an
important step is taken towards including radiation 
backreaction but the backreaction on the loop production 
function, which feeds into the loop distribution function,
is ignored.

\bibitem{Dubath:2007mf}
  F.~Dubath, J.~Polchinski and J.~V.~Rocha,
  arXiv:0711.0994 [astro-ph].

\bibitem{Vincent:1997cx}
  G.~Vincent, N.~D.~Antunes and M.~Hindmarsh,
  Phys.\ Rev.\ Lett.\  {\bf 80}, 2277 (1998)

\bibitem{Rocha:2007ni}
  J.~V.~Rocha,
  Phys.\ Rev.\ Lett.\  {\bf 100}, 071601 (2008)

\bibitem{LeeJokipii1976}
 L.~C.~Lee and J.~R.~Jokipii,
 Ap.\ J.\ {\bf 206}, 735 (1976).

\bibitem{Katz:2003et}
  C.~A.~Katz, J.~N.~Hewitt, B.~E.~Corey and C.~B.~Moore,
  arXiv:astro-ph/0304260.

\bibitem{Barr:1987xm}
  S.~M.~Barr and A.~M.~Matheson,
  Phys.\ Rev.\  D {\bf 36}, 2905 (1987).

\bibitem{Barr:1987ij}
  S.~M.~Barr and A.~M.~Matheson,
  Phys.\ Lett.\  B {\bf 198}, 146 (1987).

\bibitem{Ferrer:2006ne}
  F.~Ferrer, H.~Mathur, T.~Vachaspati and G.~D.~Starkman,
  Phys.\ Rev.\  D {\bf 74}, 025012 (2006)

\bibitem{Ostriker:1986xc}
J.~P.~Ostriker, A.~C.~Thompson and E.~Witten,
Phys.\ Lett.\  B {\bf 180}, 231 (1986).

\bibitem{Sanchez:1991jv}
N.~G.~Sanchez and M.~Signore,
Phys.\ Lett.\  B {\bf 261}, 21 (1991).

\bibitem{Damour:2000wa}
  T.~Damour and A.~Vilenkin,
  Phys.\ Rev.\ Lett.\  {\bf 85}, 3761 (2000)
%
  X.~Siemens, J.~Creighton, I.~Maor, S.~Ray Majumder, K.~Cannon and J.~Read,
  Phys.\ Rev.\  D {\bf 73}, 105001 (2006)

\bibitem{Ferrer:2005xva}
  F.~Ferrer and T.~Vachaspati,
  Phys.\ Rev.\ Lett.\  {\bf 95}, 261302 (2005)

\bibitem{Garfinkle:1987yw}
  D.~Garfinkle and T.~Vachaspati,
  Phys.\ Rev.\  D {\bf 36}, 2229 (1987).


\end{thebibliography}
\end{document}